\documentclass[superscriptaddress,aps,preprintnumbers,amsmath,amssymb,prd,nofootinbib,preprint,longbibliography]{revtex4-1}
\pdfoutput=1
\usepackage[utf8]{inputenc}
\usepackage[T1]{fontenc}
\usepackage{comment}
\usepackage{graphicx}
\usepackage{epstopdf}
\usepackage{dcolumn}
\usepackage{bm}
\usepackage[unicode=true]{hyperref}
\usepackage{color}
\usepackage{amsmath,amssymb}
\usepackage[sort&compress]{natbib}
\usepackage{float}
\usepackage{tikz}
\usepackage{tikz-cd}
 
\newcommand{\tr}{\text{tr}} 
\usepackage{colortbl}

\numberwithin{equation}{section}
\renewcommand{\thesection}{\arabic{section}}

\makeatletter
\renewcommand{\p@subsection}{}
\renewcommand{\p@subsubsection}{}
\makeatother

\begin{document}


\def\a{\alpha}
\def\b{\beta}
\def\c{\varepsilon}
\def\d{\delta}
\def\e{\epsilon}
\def\f{\phi}
\def\g{\gamma}
\def\h{\theta}
\def\k{\kappa}
\def\l{\lambda}
\def\m{\mu}
\def\n{\nu}
\def\p{\psi}
\def\q{\partial}
\def\r{\rho}
\def\s{\sigma}
\def\t{\tau}
\def\u{\upsilon}
\def\v{\varphi}
\def\w{\omega}
\def\x{\xi}
\def\y{\eta}
\def\z{\zeta}
\def\D{\Delta}
\def\G{\Gamma}
\def\H{\Theta}
\def\L{\Lambda}
\def\F{\Phi}
\def\P{\Psi}
\def\S{\Sigma}

\def\o{\over}
\def\beq{\begin{align}}
\def\eeq{\end{align}}
\newcommand{\gsim}{ \mathop{}_{\textstyle \sim}^{\textstyle >} }
\newcommand{\lsim}{ \mathop{}_{\textstyle \sim}^{\textstyle <} }
\newcommand{\vev}[1]{ \left\langle {#1} \right\rangle }
\newcommand{\bra}[1]{ \langle {#1} | }
\newcommand{\ket}[1]{ | {#1} \rangle }
\newcommand{\EV}{ {\rm eV} }
\newcommand{\KEV}{ {\rm keV} }
\newcommand{\MEV}{ {\rm MeV} }
\newcommand{\GEV}{ {\rm GeV} }
\newcommand{\TEV}{ {\rm TeV} }
\newcommand{\1}{\mbox{1}\hspace{-0.25em}\mbox{l}}
\newcommand{\headline}[1]{\noindent{\bf #1}}
\def\diag{\mathop{\rm diag}\nolimits}
\def\Spin{\mathop{\rm Spin}}
\def\SO{\mathop{\rm SO}}
\def\O{\mathop{\rm O}}
\def\SU{\mathop{\rm SU}}
\def\U{\mathop{\rm U}}
\def\Sp{\mathop{\rm Sp}}
\def\SL{\mathop{\rm SL}}
\def\tr{\mathop{\rm tr}}
\def\mpl{M_{\rm Pl}}

\def\IJMP{Int.~J.~Mod.~Phys. }
\def\MPL{Mod.~Phys.~Lett. }
\def\NP{Nucl.~Phys. }
\def\PL{Phys.~Lett. }
\def\PR{Phys.~Rev. }
\def\PRL{Phys.~Rev.~Lett. }
\def\PTP{Prog.~Theor.~Phys. }
\def\ZP{Z.~Phys. }

\def\dd{\mathrm{d}}
\def\ff{\mathrm{f}}
\def\BH{{\rm BH}}
\def\inf{{\rm inf}}
\def\ev{{\rm evap}}
\def\eq{{\rm eq}}
\def\SM{{\rm sm}}
\def\Mpl{M_{\rm Pl}}
\def\GeV{{\rm GeV}}
\newcommand{\Red}[1]{\textcolor{red}{#1}}
\newcommand{\RC}[1]{\textcolor{blue}{\bf RC: #1}}

\title{
Neutrino mass matrices 
	from localization in M-theory on $G_2$ orbifold 
}

\author{Eric Gonzalez}
\email{ericgz@umich.edu}
\affiliation{\small Leinweber Center for Theoretical Physics, Department of Physics, University of Michigan, Ann Arbor, Michigan 48109, USA}
\author{Gordon Kane}
\email{gkane@umich.edu}
\affiliation{\small Leinweber Center for Theoretical Physics, Department of Physics, University of Michigan, Ann Arbor, Michigan 48109, USA}
\author{Khoa Dang Nguyen}
\email{kdng@umich.edu}
\affiliation{\small Department of Mathematics, University of Michigan, Ann Arbor, MI 48109, USA }

\begin{abstract}
M-theory compactified on a $G_2$ manifold with resolved $E_8$ singularity is a promising candidate for a unified theory. The experimentally observed masses of quarks and charged leptons put a restriction on the moduli of the $G_2$ manifold. These moduli in turn uniquely determine the Dirac interactions of the neutrinos. In the paper, we explicitly compute the Dirac terms for neutrino mass matrix using the moduli from a localized model with resolved $E_8$ singularities on a $G_2$ manifold. This is a novel approach as the Dirac terms are not assumed but derived from the structure of quarks' and charged leptons' masses. Using known mass splittings and mixing angles of neutrinos, we show the acceptable region for Majorana terms. We also analyse the theoretical region for Majorana terms induced from the expectation values of right handed neutrinos through the Kolda-Martin mechanism. The intersection of the two regions indicates a restriction on neutrino masses. In particular, the lightest neutrino must have small but non-zero mass. Moreover, this also puts constraints on possible Majorana contributions from K\"ahler potential and superpotential, which can be traced down to a restriction on the geometry.We conclude that the masses of the two heavier light neutrinos are about $0.05 \text{ eV}$  and $0.009  \text{ eV}$  ($0.05 \text{ eV} $ and $0.05 \text{ eV} $))  for normal (inverted) hierarchy. In both hierarchies, we predict the light neutrinos are mostly Dirac type. Hence neutrino-less double-beta decay will be small. This is a testable result in a near future. Some bounds on heavy neutrinos are also derived.
\end{abstract}

\date{\today}
\maketitle
\pagebreak
\tableofcontents
\newpage
\section{Introduction}
The origin of the light left handed neutrinos in the Standard Model (SM) has been a mystery. Cosmological probes have constrained the sum of the left handed neutrino masses to be $\Sigma m_\nu < 0.12$ $(0.15)\ \text{eV}$ for normal (inverted) ordering \cite{deSalas:2020pgw}. Neutrino mass splittings observed from neutrino oscillation are $\Delta m_{12}^2= 7.6 \times 10^{-5}$ $eV^2$, and $\Delta m_{13}^2 = 2.5 \times 10^{-3}$ $eV^2$ \cite{deSalas:2020pgw}. Moreover, the oscillation angles are about $\theta_{12} = 33.44^\circ$, $\theta_{23} = 49.0^\circ$, and $\theta_{13} = 8.57^{\circ}$  \cite{deSalas:2020pgw,salas20202020}, which can be used to explicitly compute the flavor components of mass eigenstates. 

Due to none-zero mixing angles, neutrino flavor eigenstates (electron, muon, and tau) are not the same as the neutrino mass eigenstates (simply labeled “1”, “2”, and “3”). It is not known which of these three is the heaviest. In analogy with the mass hierarchy of the charged leptons, the configuration with mass 2 being lighter than mass 3 is conventionally called the “normal hierarchy”, while in the “inverted hierarchy”, the opposite would hold. Several major experimental efforts are underway to help establish which is correct. Current data favors the normal hierarchy, although the confidence for this hierarchy has been decreasing over the years \cite{deSalas:2020pgw}. In this paper we will assume the normal hierarchy first, and then apply similar framework to the inverted hierarchy.

We show that viable neutrino masses can arise within the framework of M theory with resolved $E_8$ singularities, which is a highly non-trivial result, given the constrained nature of M theory constructions.
From our previous work \cite{Gonzalez:2020nyy}, we numerically compute a local solution for moduli of $G_2$ manifold from the experimental masses of quarks and charged leptons. As these moduli locally control the geometry structure of the manifold, they determine all other interactions in the model. Therefore, we can use them to compute the Dirac terms of the neutrinos. This distinguishes our approach from previous works with neutrino Dirac mass \cite{Acharya:2016kep,Mohapatra:1999zr,Chen:2012jg,Barr:2000ka,Rodejohann:2003gz, Goh:2003hf, Damanik:2010rv, Damanik:2010xd, Akhmedov:1999tm, Ma:2021bzl} as we do not make an estimation, instead we compute the Dirac terms explicitly. 

The origin of Majorana mass terms has been complicated to realize from the string theory perspective \cite{Acharya:2016kep}. For instance, it is possible to obtain large Majorana mass terms from instanton effects \cite{Acharya:2006ia, Blumenhagen:2006xt, Ellis:2014kla}, large volume compactification \cite{Conlon:2007zza}, or orbifold compactifications of the heterotic string \cite{Buchmuller:2007zd}.  In this work, we use the Kolda-Martin mechanism \cite{Kolda:1995iw, Costa:1986jb} to generate vaccumn expectation values (VEVs) for the scalar components of right handed neutrino supermultiplets and their conjugates. The Kolda-Martin (K-M) mechanism includes effects of non-perturbative terms via the K\"ahler potential. A similar approach has been done by Acharya et al for an $SO(10)$ gauge group \cite{Acharya:2016kep}. Our work expands the idea to an explicit resolved $E_8$ singularities model, with three generations fitting the experimental data for quarks and charged leptons, and computes neutrino Dirac terms. The computed Dirac terms put constraints on the Majorana terms through the see-saw mechanism, and the Majorana terms are generated from the VEVs of the conjugates of right handed neutrinos.

Additionally, when the right handed neutrinos get VEVs, we inevitably generate bilinear R-parity violating terms of the form $\epsilon_{ij} L_i H_j$. There are many works dedicated to study these terms \cite{Barbier:2004ez, Cohen:2019cge, Diaz:1997ef}. In general, due to the presence of large Majorana terms, the bilinear mixing between Higgs and leptons may spoil the Higgs physics. It is more favorable to have a small $\epsilon_{ij}$. This puts a stringent constraints on the aforementioned VEVs. In this paper, we show that there are solutions for the VEVs in which the mixing between leptons and Higgses is minimal. As a result of the constraints, with a generic un-suppressed K\"ahler potential coefficient, the lightest neutrino can be neither massless nor heavy.

Futhermore, the nature of lightest neutrinos are expected to be determined in a near future. The most important process for this effort is the neutrinoless double-beta decay, in which the total lepton number is violated by two units. The prediction for the Standard Model including non-perturbative effects via K-M mechanism is that the neutrinos are Dirac. That implies neutrinoless double beta decay will be small under those assumptions, and therefore a good window for new physics, e.g.neutralinos (which are Majorana particles), R-parity violating interactions, and new physics in general. If the light neutrinos are significantly Majorana, the experiments should be able to detect them. Otherwise, the particles must be mostly Dirac \cite{Oberauer}. In this paper, we predict that the light particles are mostly Dirac, hence the decay will be small.

This paper is organized as following: section \ref{background} will briefly cover the local model of M theory compactified on a $G_2$ manifold with resolved $E_8$ singularities \cite{Gonzalez:2020nyy}. Section \ref{terms} will list all of the contributions to the neutrino mass matrix. Section \ref{VEVs} discusses the VEVs for the right handed neutrinos and their conjugates through the K-M mechanism while discussing the $\epsilon_{ij}$ problem. Section \ref{matrix} contains the computed Dirac matrix and sets up the framework for the neutrino mass matrix. In section \ref{majorana} we discuss the Majorana mass matrix from the experimental data and from the right handed neutrino VEVs. In section \ref{limit} we deduce a limit on the neutrino masses. We predict the masses of the mass eigenstate neutrinos, though we cannot yet exclude one of the normal or inverted case. This will lead to section \ref{Dirac-Majo ratio} where the ratio of Dirac to Majorana components are estimated. Finally, some insight about heavy neutrino masses are presented in section \ref{heavy mass}.
\section{Background}\label{background}
\subsection{General setup}
M-theory is compactified on a $G_2$ manifold which is a 7-d manifold. We are interested in $G_2$ manifold as an ALE fiberation of $\mathbb{C}^2/\Gamma$ on a 3-d base $M_3$. Locally the manifold looks like
\begin{align}
M_3 \times \widehat{\mathbb{C}^2/\Gamma}
\end{align}
where $\widehat{\mathbb{C}^2/\Gamma}$ is some resolution of ADE singular space $\mathbb{C}^2/\Gamma$ which is a quotient of complex space $\mathbb{C}^2$ by a finite subgroup $\Gamma$ of $SU(2)$, resulting in a singularity at the origin. In our case, the singularity is $E_8$.

There is a metric $g$, 3-form $C$, gravitino spinor $\Psi$, and $G_2$ structure 3-form $\omega$ on the 7-d $X$ . The 3-form $\omega$ completely determines the metric $g$ of the $G_2$ manifold; therefore, $C$ ,$\Psi$ and $\omega$ govern all of the physics in M-theory. Integrating $C$ and $\omega$ on the basis of three cycles of $X$ gives axions $a_i$ and moduli fields $s_i$ respectively \cite{Acharya:2007rc,Acharya:2008zi,Acharya:2004qe} . The superfield $\Phi_i$ is then of the form
\begin{align}
	\Phi_i=a_i + is_i + \text{fermionic terms}
\end{align}

The moduli $s_i$ controls the size of the three cycles in $X$. When $s_i$ varies along the base $M_3$, the singularity $E_8$ can be deformed and resolved to lower singularities or even completely smooth points. These singularities have a one-to-one correspondence with the gauge group of M theory \cite{Witten:2001uq}. The moduli $s_i$ corresponding to an $E_8$ singularity are listed in \cite{Bourjaily:2007vx}. Vanishing moduli correspond to the simple roots of the gauge group. When a simple root modulus vanishes the gauge group is enhanced to a larger one. Furthermore, chiral fermions localize in places along these ADE singularites where non-simple-roots three cycles vanish to form conical singularities \cite{Acharya:2001gy}. Bourjaily et al \cite{Bourjaily:2007vx} and Gonzalez et al \cite{Gonzalez:2020nyy} give detailed explanations and example computations for the presentations of the chiral fermions.  In our model, the resolved $E_8$ singularity results in matter as in Table 9 and Table 10 in \cite{Bourjaily:2007vx}. In the following, the charges are listed in the same order as in Table 10 in \cite{Bourjaily:2007vx}, namely in order $a,b,c,d,Y$. Then we have
\begin{align}
	&E_8 \rightarrow SU(3) \times SU(2) \times U(1)_a \times U(1)_b \times U(1)_c \times U(1)_d \times U(1)_Y.
\end{align}
The hypercharge Y has a factor of $6$ compared to the conventional hypercharge normalization to make all the charges integer and does not effect the calculation. The charge $a$ is identically set to zero, following \cite{Gonzalez:2020nyy}, to drop terms dependent on a. Then, the relevant particles for this study are

\begin{align}
&H^d_1=({\bf 1} ,{\bf 1})_{(1, 1,2, 2,-3)}& &H^u_1=({\bf 1} ,{\bf 1})_{(1, 1,2, 2,3)}&	&L_1= ({\bf 1} ,{\bf 2})_{( 1, 1, -1,3,-3)}&	&\nu_1^c= ({\bf 1} ,{\bf 1})_{( 1, 1, -1,-5,0)} \nonumber \\
&H^d_2=({\bf 1} ,{\bf 1})_{(1,-1,2, 2,-3)}& &H^u_2=({\bf 1} ,{\bf 1})_{(1,-1,2, 2,3)}&	&L_2= ({\bf 1} ,{\bf 2})_{( 1, -1, -1,3,-3)}&	&\nu_2^c= ({\bf 1} ,{\bf 1})_{( 1, -1, -1,-5,0)}\nonumber\\
&H^d_3=({\bf 1} ,{\bf 1})_{(0,-1,2, 2,-3)}& &H^u_3=({\bf 1} ,{\bf 1})_{(0,-1,2, 2,3)}&	&L_3= ({\bf 1} ,{\bf 2})_{(-2, 0, -1,3,-3)}&	&\nu_3^c= ({\bf 1} ,{\bf 1})_{(-2, 0, -1,-5,0)}	.
\end{align}

 The reason for $a=0$ is to allow large top quark mass \cite{Gonzalez:2020nyy}. Note that the simple root cycles do not shrink under this condition, so there is no enhanced gauge group. This is similar to taking the diagonal $U(1)_a \times U(1)_b$. Notice that a $\mu$ term  $H^u_i H^d_j$ is generally not allowed, but can be generated by the Giudice-Masiero mechanism \cite{Chen:2012jg}.

\subsection{Yukawa couplings}
The couplings for the interactions in the superpotential are given by the instanton effect \cite{Braun:2018vhk, Atiyah:2001qf, Acharya:2007rc, Hubner:2020yde, Pantev:2009de}
\begin{align}\label{Yukawa}
	Y = \frac{1}{\Lambda} e^{-V_{\text{3 cycles}}}
\end{align}
where $\Lambda$ is a scaling factor proportional to the volume of the $G_2$ manifold \cite{Acharya:2007rc, Braun:2018vhk}. In our model, the local moduli are not enough to determine the volume, so we treat $\Lambda$ as a parameter. $V_{\text{3 cycles}}$ is the volume of the three cycles stretching between the three singularities where the three particles in the cubic terms are located. This volume is a function of the moduli
\begin{align}
    Vol(\Sigma_{ABC})=\frac{1}{2} (-v_A^T H_A^{-1} v_A - v_B^T H_B^{-1} v_B 
&+ (v_A + v_B)^T (H_A + H_B)^{-1} (v_A+ v_B)).
\end{align}
Here, $\Sigma_{ABC}$ is a three cycle covering three particle singularities A, B, and C. Moreover, each singularity's location on $M_3$ is determined by the critical point of
\begin{align}
    f=\frac{1}{2} t^T H t + v^T t + c
\end{align}
where $t$ is the local 3-d coordinate on $M_3$, $H$ is a $3 \times 3$ matrix, $v$ is a 3-vector, and c is a scalar. Using this setup, we can write down the mass matrix for quarks and charged leptons. Then, by fitting to experimental data, we can find the solutions for $f_i$'s in the local model. We will use the fit result of b, c, d, and Y from \cite{Gonzalez:2020nyy}. In a full theory on a determined $G_2$ manifold, the moduli should uniquely determine every other quantity in the theory as they determine the geometry of the manifold. In our local model, as there is some global structure we are missing, the $f_i$'s will determine many quantities, such as Dirac neutrino terms, but leave some other quantities, such as Majorana terms and the soft breaking mechanism \cite{Acharya:2016kep}, subject to tuning. Nonetheless, most of our main results will not depend of the tuning.

\section{Terms}\label{terms}

\subsection{Neutrino-neutrino mixing terms}
At tree level, the contribution from the superpotential is
\begin{align}\label{Dirac_contribution}
	W_{tree} \supset  y_{123} H^u_1 L_2 \nu^c_3  +  y_{132} H^u_1 L_3 \nu^c_2 + y_{312} H^u_3 L_1 \nu^c_2 +  y_{321} H^u_3 L_2 \nu^c_1\\
	+  y_{213} H^u_2 L_1 \nu^c_3 + y_{231} H^u_2 L_3 \nu^c_1 + y_{333} H^u_3 L_3 \nu^c_3	
\end{align}
where $y_{ijk}$ are coupling constants computed from Eq.~\ref{Yukawa}. There are also contributions to the same terms from the K\"ahler potential with coefficients of order $\frac{1}{m_{pl}}$ which is negligible \cite{Acharya:2016kep}. Similar to the work done by Acharya et al \cite{Acharya:2016kep} to generate a Majorana mass term, we get contributions to the superpotential of the form
\begin{align}\label{Majorana_contribution}
	W \supset  \sum_{0 \le h,l,m\le n}\sum_{i, j,k=1,2,3} \frac{C_{h,l,m}}{m_{pl}^{2n-3}} (\nu_i^c \bar{\nu_i^c})^{h}(\nu_j^c \bar{\nu_j^c})^l(\nu_k^c \bar{\nu_k^c})^m
\end{align}
where $m_{pl}=2.4 \times 10^{18}$ GeV is the reduced Planck mass. Ideally, the constants $C_{h,k,l}$ should be determined completely by the moduli of the $G_2$ manifold. In the local model of \cite{Gonzalez:2020nyy}, we do not consider $\bar{\nu}_j$ fields as they are controlled by global moduli beyond the local patch. As a result, $C_{h,l,m}$ is considered a tunable parameter in our local model.

Contributions from the K\"{a}hler potential to the same terms are expected. They can be computed from the full  K\"{a}hler potential \cite{Beasley:2002db, Lukas_2004}
\begin{align}
    K =-3 \log\big(\frac{V}{2\pi}\big)
\end{align}
where $V$ is  the volume of $G_2$ manifold. Unfortunately, the precise dependence of the volume on the global moduli in resloved $E_8$ orbifold is unknown. We assume it is not significant due to the generic suppression as in \cite{Acharya:2016kep}. 

By solving D term and F term equations from the terms in equation (\ref{Majorana_contribution}), one can find the VEVs for right handed neutrinos. Assuming the leading term is quartic which we will justify later, the Majorana mass terms in the superpotential would have the form
\begin{align}\label{Majorana mass term}
   \sum_{i,j} \frac{C_{2,1}}{m_{pl}}( \langle \bar{\nu_i^c} \rangle \langle \bar{\nu_j^c} \rangle)  \nu_i^c  \nu_j^c .
\end{align}
Additionally, we also receive terms of the form $L_i H^u_j$ from expression (\ref{Dirac_contribution}) when right handed neutrinos get VEVs. We will discuss this in section \ref{VEVs}. In the same manner, the Dirac mass terms emerge from Eq.~\ref{Dirac_contribution} when the Higgses get VEVs.

\subsection{Mixing Matter with Higgs Superfields}
When the scalar components of the right handed neutrino superfields $\nu^c_i$ get VEVs, cubic terms of the form $Y_{ijk} H^u_i L_j \nu^c_k$ will give rise to the mixing between $L_j$ and $H^u_i$  superfields. They appear in superpotential as

\begin{align}
	\mu_{ij} H^u_i L_j
\end{align}
where 
\begin{align}
    \mu_{ij}= Y_{ijk} \langle \nu^c_k \rangle.
\end{align}
This mixing can potentially spoil the Higgs physics, so it is generally more favorable to consider small $\mu_{ij}$ relative to Dirac mass terms in the neutrino mass matrix. This creates a stringent condition which requires $\langle \nu^c_k \rangle < \langle H^u_i \rangle$ while $\langle \bar{\nu^c_k} \rangle$ remains large due to Eq.~\ref{Majorana mass term} and the see-saw mechanism. This will be realized in section \ref{VEVs}. 

Furthermore, the presence of R-parity violating bilinear terms (B-RPV) induces a sub-electroweak scale (EWS) VEV on the scalar components of the $\nu$-type fields. In our case, below the EWS, we expect all $\nu$-type scalars to acquire a non-vanishing VEV, generating a mixing between right handed neutrino and Higgsinos \cite{Acharya:2016kep}
\begin{align}
	\epsilon_{ij} H^{u}_i \nu^c_j 
\end{align}
Although this can create some correction to our analysis, the contribution is usually expected to be smaller then the Dirac mass terms \cite{Acharya:2016kep}.

\subsection{Mixing Matter with Gauginos}
Finally, as in the Minimal Supersymmetric Standard Model (MSSM), the presence of VEVs will mix some fermions with gauginos through kinetic terms, namely the Higgsinos with $\tilde{B}_1, \tilde{W}_0$ due to the Higgses VEVs \cite{Binetruy}. In our case we also have $\nu^c$-type and $\nu$-type scalar VEVs, which will mix gauginos with matter fermions through kinetic terms. Explicitly, we have, for the $SU(2)$ states (left-handed neutrinos),
\begin{align}
L \supset	g' \tilde{B} \langle \tilde{\nu_i} \rangle \nu_i + g \tilde{W}^0 \langle \tilde{\nu_i} \rangle \nu_i + g_b  \tilde{B}_b \langle \tilde{\nu_i} \rangle \nu_i + g_c  \tilde{B}_c \langle \tilde{\nu_i} \rangle \nu_i  + g_d  \tilde{B}_d \langle \tilde{\nu_i} \rangle \nu_i 
\end{align}
where the coefficients are gauge couplings. There will be an extra (charge $\times \sqrt(2)$) coefficient for each specific particle \cite{Binetruy}. For the $\nu^c$-states, which are singlets under the SM gauge group, mixing takes the form
\begin{align}
L \supset	g_b  \tilde{B}_b \langle \tilde{\bar{\nu_i}} \rangle \bar{\nu_i} + g_c  \tilde{B}_c \langle \tilde{\bar{\nu_i}} \rangle \bar{\nu_i}  + g_d  \tilde{B}_d \langle \tilde{\bar{\nu_i}} \rangle \bar{\nu_i} .
\end{align}

\subsection{General Mass Matrix}
Combining all of the previous arguments, we can write down the general mass matrix for neutrinos. Considering the basis 
\begin{align}
	(\tilde{B},\tilde{W}^0,\tilde{B}_{b,c,d},H^{u0}_{1,2,3}, \nu_{1,2,3},\nu_{1,2,3}^c),
\end{align}

the mass matrix will be 
\begin{align}
	M = \begin{pmatrix}
	M_{\chi^0}^{8 \times 8} & M^{8 \times 6}_{\chi \nu}\\
	(M^{8 \times 6}_{\chi \nu})^T &  M^{6 \times 6}_{\nu} \\
	\end{pmatrix}.
\end{align}
where $M^{8 \times 6}_{\chi \nu}$ is the mixing sub-matrix between gauginos, Higginos, and neutrinos which is insignificant in our analysis of the magnitude of neutrino masses. $M_{\chi^0}^{8 \times 8}$ is the pure gauginos-Higginos sub-matrix. Although this sub-matrix can be significant in size, the small mixing with neutrinos makes $M_{\chi^0}^{8 \times 8}$ irrelevant for the magnitude of neutrino masses. It would be interesting to study their effects in detail in future works. Thus, for the scope of this paper, we will focus only on the neutrino sub-matrix $M^{6 \times 6}_{\nu}$.

\section{VEVs of right handed neutrinos and their conjugates} \label{VEVs}
In order to explicitly write down the entries for $M^{6 \times 6}_{\nu}$, in this section we will consider a semi-general method to give VEVs to right handed neutrinos and their conjugates.
\subsection{Case 1: No Mixing}
	First, we consider a standard superpotential that gives rise to right-handed neutrino VEVs without mixing of families
\begin{align}
\mu \nu_i^c \bar{\nu}_i^c + \frac{C_{n,0,0}}{m_{pl}^{2n-3}}( \nu_i^c \bar{\nu^c}_i)^n
\end{align}
where $\mu= m_{3/2} \frac{s}{m_{pl}}=\mathcal{O}(10^3)$ GeV with $m_{3/2}=\mathcal{O}(10^4)$ GeV is the mass of gravitino, $\frac{s}{m_{pl}} \equiv 0.1 $ GeV is a generic moduli VEVs contribution in K\"ahler potential \cite{Acharya:2016kep}. The latter should be determined completely from the value of the moduli if we have a complete description of $G_2$ manifold. Unfortunately, we will use this estimated value due to our lack of knowledge for a complete $G_2$ structure.

 D-flat directions implies
 \begin{align}\label{D}
 \sum_i q^j_i\big(|\langle \nu_i^c  \rangle|^2 - |\langle \bar{\nu}_i^c \rangle|^2\big)- \xi_j=0
 \end{align}
 for $j=b,c,d, Y$ and $\xi $'s are  from Fayet–Iliopoulos terms. F-flat directions  give
  \begin{align}
     \mu \nu_i^c + \frac{n C_{n,0,0}}{m_{pl}^{2n-3}}( \nu_i^c)^n (\bar{\nu^c}_i)^{n-1}=0\\
     \mu \bar{\nu}_i^c + \frac{n C_{n,0,0}}{m_{pl}^{2n-3}}( \nu_i^c)^{n-1}(\bar{\nu}^c_i)^n=0
 \end{align}
 The VEVs for $\nu_i^c$ can be problematic because they can create terms such as $y \langle \nu^c\rangle H^u L$ which may spoil Higgs physics. On the other hand, large VEVs for $\bar{\nu_i}^c$ are needed to generate large Majorana terms for right handed neutrinos and hence see-saw mechanism. Thus, we consider $\langle \nu_i^c  \rangle = \epsilon_i \langle \bar{\nu}_i^c \rangle$. From F-terms, this will imply
\begin{align}
\langle \nu_i^c  \rangle = \epsilon_i \langle \bar{\nu}_i^c \rangle =\sqrt{\epsilon_i} \big(-\frac{\mu m_{pl}^{2n-3}}{n C_{n,0,0}}\big)^{\frac{1}{2(n-1)}}.
\end{align}
Plugging this into the D-term, we get a restriction for Fayet–Iliopoulos coefficients.
\begin{align}
    \xi_b &= (\epsilon_1^2-1)  \langle \bar{\nu}_1^c \rangle - (\epsilon_2^2-1)   \langle \bar{\nu}_2^c \rangle \\
    \xi_c &= -\sum_{i=1}^3(\epsilon_i^2-1)  \langle \bar{\nu}_i^c \rangle\\
    \xi_d &= -5\sum_{i=1}^3(\epsilon_i^2-1) \langle \bar{\nu}_i^c \rangle\\
    \xi_Y &=0
\end{align}

This cannot give too much texture to Majorana terms without tuning $ C_{n,0,0}$. From the observed data, as we will see later, a rich texture is needed. Therefore, it is inviting to consider the mixing case.
\subsection{Case 2: Mixing with Two Families}
Consider the simplest mixing K\"aler potential
\begin{align}
	\mu \nu_i \bar{\nu}_i + \mu \nu_j \bar{\nu}_j + \frac{
	C_{n-k,k,0}}{m_{pl}^{2n-3}}( \nu_i \bar{\nu}_i)^{n-k} ( \nu_j \bar{\nu}_j)^{k}.
\end{align}
The D-flat equations are the same as in Eq.~\ref{D}. Again we consider $\langle \nu_i^c  \rangle = \epsilon_i \langle \bar{\nu}_i^c \rangle$. F-flat directions give
\begin{align}
    	\mu \nu_i + (n-k) \frac{C_{n-k,k,0}}{m_{pl}^{2n-3}}( \nu_i)^{n-k} ( \nu_j \bar{\nu}_j)^{k} (\bar{\nu}_i)^{n-k-1}=0,\\
    	\mu \nu_j + (k) \frac{C_{n-k,k,0}}{m_{pl}^{2n-3}}( \nu_i \bar{\nu}_i)^{n-k} ( \nu_j)^{k} (\bar{\nu}_j)^{k-1}=0\\
    	\text{Interchange } \nu \leftrightarrow \bar{\nu}.
\end{align}
which imply
\begin{align}
    \langle \nu_i^c \rangle= \epsilon_i  \langle \bar{\nu}_i^c \rangle=\sqrt{\epsilon_i}\Bigg[ -\frac{\mu}{C_{n-k,k,0}} \frac{(n-k)^{k-1}}{k^k} m_{pl}^{2n-3}\Bigg]^{\frac{1}{2(n-1)}},\\
   \langle  \nu_j^c \rangle = \epsilon_j\langle \bar{\nu}_j^c \rangle =\sqrt{\epsilon_j} \Bigg[ -\frac{\mu}{C_{n-k,k,0}} \frac{k^{n-k-1}}{(n-k)^{n-k}} m_{pl}^{2n-3}\Bigg]]^{\frac{1}{2(n-1)}}.
\end{align}
A hierarchy for Majorana terms is possible here as right handed anti-neutrinos from different families get different VEVs.  
\subsection{Case 3: Mixing with Three Families}
We can consider the simplest mixing of three families in the K\"ahler potential
\begin{align}
    	\mu \nu_1^c \bar{\nu}_1^c + \mu \nu_2^c \bar{\nu}_2^c +  \mu \nu_3 \bar{\nu}_3 + \frac{C_{h,k,l}}{m_{pl}^{2n-3}}( \nu_1 \bar{\nu}_1)^{h} ( \nu_2 \bar{\nu}_2)^{k}  ( \nu_3 \bar{\nu}_3)^{l}.
\end{align}
where $h+k+l=n$. The D-flat equations are the same as in  Eq.~\ref{D}. Again we consider $\langle \nu_i^c  \rangle = \epsilon_i \langle \bar{\nu}_i^c \rangle$. Then, F-term equations are
\begin{align}
	\mu +  \frac{h C_{h,k,l}}{m_{pl}^{2n-3}}( \nu_1^c \bar{\nu}_1^c)^{h-1} ( \nu_2^c \bar{\nu}_2^c)^{k}  ( \nu_3 \bar{\nu}_3^c)^{l}=0,\\
	\text{Permute 3 pairs }  (1,h), (2,k) \text{, and } (3,l),\\
	\text{permute } \nu \leftrightarrow \bar{\nu}.
\end{align}
The solution is
\begin{align}\label{VEVs3}
    \langle \nu_1^c \rangle =  \epsilon_i \langle \bar{\nu}_i^c \rangle= \sqrt{\epsilon_i}\Bigg[-\frac{\mu h^{k+l+1} m_{pl}^{2n-3}}{C_{h,k,l} k^k l^l}\Bigg]^\frac{1}{2(n-1)},\\
    \text{Permute 3 pairs } (1,h), (2,k) \text{, and } (3,l).
\end{align}

Note that in all of the above cases, in practice, we can drop the negative signs inside the brackets as they can be absorbed as a phase in the oscillation matrix of neutrinos. Another scenario is that one of the right handed neutrinos completely decouples from the other two. The K\"ahler potential will then be a sum of case 1 and case 2, and the solutions are the same as case 1 and case 2.

\section{Mass Matrix from Neutrino Mixing}\label{matrix}
\subsection{Mass Matrix Setup}
	We investigate the matrix with only right handed neutrinos and left handed neutrinos. Using the moduli values computed from quarks and charged lepton mass in \cite{Gonzalez:2020nyy} \footnote{Note that although we can only find one solution in \cite{Gonzalez:2020nyy}, it is likely not unique. Study about the uniqueness of local solution is left for future study.}, we compute Dirac mass terms from the cubic yukawa couplings at tree level
	\begin{align}
	W_{tree} \supset  y_{123} H^u_1 L_2 \nu^c_3  +  y_{132} H^u_1 L_3 \nu^c_2 + y_{312} H^u_3 L_1 \nu^c_2 +  y_{321} H^u_3 L_2 \nu^c_1\\
	+  y_{213} H^u_2 L_1 \nu^c_3 + y_{231} H^u_2 L_3 \nu^c_1 + y_{333} H^u_3 L_3 \nu^c_3
	\end{align}
	where $y_{ijk}$'s are computed from the moduli. The Yukawa couplings $y_{ijk}$ form a matrix
\begin{align}
    Y=\begin{pmatrix}
	&0 &6.93 \times 10^{-7} &4.52 \times 10^{-10}\\
	&7.25\times 10^{-1} &0 &3.19 \times 10^{-1}\\
	&2.53 \times 10^{-5} &1.71 \times 10^{-2} &3.22\times 10^{-2}
	\end{pmatrix}.
\end{align}
When the Higgs get VEVs, the Dirac terms (in GeV) are approximately
\begin{align}
    D=\begin{pmatrix}
	&0 & 2.32 \times 10^{-5} & -3.28 \times 10^{-8}\\
	&2.42\times 10^{1} &0 &-4.93 \times 10^{1}\\
	&-1.83 \times 10^{-3} & -2.64 \times 10^{0} & 1.08 \times 10^{0}
	\end{pmatrix}.
\end{align}

The first two diagonal entries vanish because there are no charge invariant terms for those. This comes down to the fact that when breaking from $E_8$, particles from the same family have the same $b$ charge. If their charges are non-zero, they cannot couple in cubic level, which is the case for the first two families with $b$ charge $\pm 1$. The explanation for the size of the rest is complicated as the Yukawa is related to the moduli by exponentiated inverse matrices. However, the significant different in sizes of the entries can be traced back to the hierarchy of the up-type quarks whose b and c charges are the same as the neutrinos.

The Majorana contribution comes form the superpotential 
	\begin{align}\label{superpotential_vevs}
		W \supset y \nu_i^c \bar{\nu_i^c} \nu_j \bar{\nu_j^c}
	\end{align}
	which was discussed in Sec.~\ref{terms}. When neutrino conjugate terms $\bar{\nu_i^c}$ get VEVs, terms of the form in equation \ref{superpotential_vevs} constitute the Majorana mass matrix $RM$. The mass matrix is in the basis of $\{L_1,L_2, L_3, \nu^c_1, \nu^c_2, \nu^c_3\}$ 
	\[\begin{pmatrix}
	& 0 &D\\
	&D^\intercal & RM
	\end{pmatrix}\]
	where $RM$ is the right-handed Majorana matrix. Notice that $RM$ must be symmetric. $RM$ gets large entries when right-handed neutrinos get VEVs. Before computing the VEVs for right-handed neutrinos through a variety of methods, we want to see if it is possible to get a sensible left-handed neutrino hierarchy and flavor-ratio for the mass eigenstates. According to the experimental data, orthonormal eigenvectors are approximately
	\begin{align}
	    V\equiv\begin{pmatrix}
	&v_1 &v_2 &v_3
	\end{pmatrix} =
	\begin{pmatrix}
	& c_{13} c_{12} & c_{13} s_{12} & s_{13}\\
	& -c_{23} s_{12}-s_{13}s_{23}c_{12} &c_{23}c_{12} -s_{13}s_{23}s_{12} & c_{13} s_{23}\\
	& s_{23}s_{12} -s_{13} c_{23} c_{12} &-s_{23}c_{12}-s_{13}c_{23} s_{12} & c_{13} c_{23} 
	\end{pmatrix}
	\end{align}
	where $c_{ij}=\cos(\theta_{ij})$, $s_{ij}=\sin(\theta_{ij})$, and  we omitted the possible phase for simplicity. We use the oscillation angles
    \begin{align}
        &\theta_{12}=33.44^\circ & &\theta_{13}=8.57^\circ&  &\theta_{23}= 49.0^\circ.
    \end{align}
	Assuming normal hierarchy, the eigenvalues are 
	\begin{align}
	    \Lambda \equiv \diag( m_1, m_2, m_3)= \diag( x, \sqrt{\Delta m_{21}^2  +x^2}, \sqrt{\Delta m_{31}^2+ x^2},)
	\end{align}
	where $x$ is the mass of the lightest left-handed neutrino and the mass-square differences are
	\begin{align}
	&\Delta m_{31}^2 = 2.32 \times 10^{-21} \text{ GeV}^2& & \Delta m_{21}^2 = 7.6 \times 10^{-23} \text{ GeV}^2
	\end{align}
	Finally, we denote the remaining components of the left-handed neutrino eigenvectors as
	\begin{align}
	  E \equiv
	\begin{pmatrix}
	&\epsilon_1 &\epsilon_2 &\epsilon_3
	\end{pmatrix}
	\end{align}

	which we expect to be small but non-zero. The final eigenvector expression is
\begin{align}\label{eigen}
    \begin{pmatrix} 
	& 0 &D\\
	&D^\intercal & RM
	\end{pmatrix}
	\begin{pmatrix}
	&V\\
	&E
	\end{pmatrix}=\begin{pmatrix}
	&V\\
	&E
	\end{pmatrix} \Lambda
\end{align}
\section{Majorana Mass Matrix}\label{majorana}
\subsection{Majorana Mass Matrix from See-Saw Mechanism}
Performing the explicit multiplication in Eq.~\ref{eigen}, we get
\begin{align}\label{See-Saw}
DE=V\Lambda \implies E = D^{-1} V \Lambda,\\
D^\intercal V + RM E = E \Lambda \implies RM D^{-1} V \Lambda = E \Lambda  - D^\intercal.\label{why RM}
\end{align}
The lightest neutrino cannot be massless, otherwise $(RM D^{-1} V -E)\Lambda$ would have a vanishing third column while $D^\intercal$ does not. Thus, $\Lambda$ is invertible. Combining the two equations we get an expression for $RM$
\begin{align}\label{RM}
RM =D^{-1} V \Lambda V^{-1} D -D^\intercal V \Lambda^{-1} V^{-1} D.
\end{align}
Notice that as $\Lambda$ has very small diagonal entries, the second term is dominant
\begin{align}\label{RM_dom}
RM \approx -D^\intercal V \Lambda^{-1} V^{-1} D.
\end{align}
For convenience, we absorb negative signs by a phase in $V$. We can investigate the small $x$ regime by writing
\begin{align}
    RM_{ij} \approx (D^\intercal V)_{i1} \frac{1}{x} (V^{-1} D)_{1j}=\frac{(V^{-1} D)_{1i}(V^{-1} D)_{1j}}{x}.
\end{align}
Thus, at small $x$, the Majorana terms will behave as a hyperbolic curve with respect to the lightest neutrino mass $x$, and the texture of $RM$, modulo the magnitude of $x$, is given by the first column of $V^{-1} D$ which is fixed.

When $m_1$ is close to the largest mass splitting, all $m_i$ have the same magnitude and the approximation becomes
\begin{align}
    RM_{ij} \approx \sum_k (D^\intercal V)_{ik} \frac{1}{m_k} (V^{-1} D)_{kj}=\frac{\sum_k (V^{-1} D)_{ki}(V^{-1} D)_{kj}}{x}
\end{align}
which is also a hyperbola with respect to $x$, although the texture of $RM$ relies on all of $V^{-1} D$ here.

To build an intuition on the magnitude of $RM$, we plug in $x=10^{-11.5} GeV$ which is about the size of the second mass splitting. The diagonalized left handed neutrino mass matrix is $\diag(4.9 \times 10^{-5}, 8.6 \times 10^{-6}, 3.2 \times 10^{-12})$, absorbing negative signs by a phase in $V$,  we get
\begin{align}
    RM= 
\begin{pmatrix}
&6.6 \times 10^{13} & 4.6 \times 10^{12} & 1.4 \times  10^{14}\\
&4.6 \times 10^{12} & 5.8 \times 10^{11} & 9.5 \times 10^{12} \\
&1.4 \times 10^{14} & 9.5 \times 10^{12} & 2.8 \times 10^{14} 
\end{pmatrix}
\end{align}

which is a symmetric matrix as we wanted. We will see that this matrix can be constructed with appropriate right-handed neutrino VEVs. For readability, the above entries of this Majorana matrix are being rounded from the actually precise values needed for the hierarchy. In fact, the hierarchy and oscillation of left-handed neutrinos can only be achieved with a high level of precision in the entries of $RM$. We cannot round the entries up because that would destroy the final hierarchy and oscillation. This is a consequence of Eq.~\ref{why RM}, where the entries of $RM$ are in general much larger than those of $\Lambda$, independent of $E$. So for the equality in Eq.\ref{why RM} to happen, entries of $RM$ need to cancel out in $RM E$ precisely to very small non-zero numbers.
\subsection{Majorana mass from VEVs of $\nu^c_i$}
We will argue that contributions beyond the order of equation \ref{Majorana mass term} will be insignificant. In fact, the contribution from order $2N$ in the superpotential is
\begin{align}
    \sum_{i,j} \frac{C_{N,N,0}}{m_{pl}^{4N-3}}\langle \bar{\nu_i^c} \rangle^{N} \langle \bar{\nu_j^c} \rangle^{N}  \langle \nu_i^c \rangle^{N-1} \langle \nu_j^c \rangle^{N-1} \nu_i^c  \nu_j^c 
\end{align}
Plugging in the VEVs from equation \ref{VEVs3}, the coefficients are of the form
\begin{align}\label{general}
\frac{C_{N,N,0} m_{pl}^{\frac{n-2N}{n-1}}}{\epsilon} \Bigg[ (hk)^{l+1} h^{k-h} k^{h-k} \frac{\mu^2}{C_n^2}\Bigg]^{\frac{2N-1}{2(n-1)}}
\end{align}
where $h, k ,l$ are permuted to get other terms. Instead of separate $n_i$ and $n_j$ for $\langle \nu_i^c \rangle$ and $\langle \nu_j \rangle$, we can consider $n_i = n_j = n$ for some fractional $n$. Assume $\frac{C_{N,1}}{C_{N+1,1}} \approx \mathcal{O}(1)$. The $h,k,l$ dependent part is also approximately $\mathcal{O}(1)$.and the coefficient is decreasing with respect to $N$ if 
\begin{align}
    \frac{\mu}{C_{N,N,0} m_{pl}}<1
\end{align}
which implies
\begin{align}
    C_{N,N,0} > \frac{\mu}{m_{pl}} \approx \frac{10^3}{10^{18}} = 10^{-15} \text{ Gev}.
\end{align}
Thus, as long as the suppression coefficient is not too small, the main contribution is always at quadric order. Henceforth, we assume $C \in [10^{-15}, 1]$ which is consistent with Acharya et al \cite{Acharya:2016kep}.
\section{Limit for Neutrinos}\label{limit}
\subsection{Lower Bound for $\epsilon_i$}
When the right handed neutrinos get VEVs, along with familiar Dirac terms of the form
\begin{align}
  y \langle H^u_j \rangle L_i \nu^c_k  ,  
\end{align}
there are terms of the from
\begin{align}
y \langle \nu^c_k \rangle L_i H^u_j.
\end{align}
which may potentially spoil the Higgs' physics. Therefore, it is desirable for the couplings to be smaller than those of the $\mu$ terms $\mu H^u_i H^d_j$ (generated at electroweak scale) and the Dirac terms. As our computed Dirac coupling $y$ is $\mathcal{O}(10)$, it is sufficient to have the right handed neutrino VEVs smaller than those of the Higgses
\begin{align}
    \langle \nu^c_i \rangle \lsim 10^2.
\end{align}
Plugging the result from \ref{VEVs3} in, we get
\begin{align}
    \sqrt{\epsilon_i} \lsim 10^2 \Bigg[\frac{\mu h^{k+l+1} m_{pl}^{2n-3}}{C_{h,k,l} k^k l^l}\Bigg]^\frac{-1}{2(n-1)} 
\end{align}
which implies
\begin{align}
    \sqrt{\epsilon_i} \lsim 10^2 \Bigg[\frac{\mu  m_{pl}^{2n-3}}{C_{h,k,l}  }\Bigg]^\frac{-1}{2(n-1)} 
\end{align}
where we have again assumed the $k,h,l$ dependent factor to be approximately $\mathcal{O}(1)$.
\subsection{Normal Hierarchy Analysis}
Using the upper bound for $\epsilon$ we can find a lower bound for the Majorana mass term
\begin{align}\label{RM upper bound}
    RM_{ij}=\frac{C_{1,1}}{m_{pl}} \langle \bar{\nu}^c_i \rangle \langle \bar{\nu}^c_j \rangle = \frac{C_{1,1}m_{pl}^{\frac{n_{ij}-2}{n_{ij}-1}} \mu^{\frac{1}{n_{ij}-1}}}{\epsilon C_{h,k,l}^{\frac{1}{n_{ij}-1}}} \ge 10^4 \Bigg[\frac{\mu  m_{pl}^{2n_{ij}-3}}{C_{h,k,l}  }\Bigg]^\frac{1}{(n_{ij}-1)} \frac{C_{1,1}m_{pl}^{\frac{n_{ij}-2}{n_{ij}-1}} \mu^{\frac{1}{n_{ij}-1}}}{ C_{h,k,l}^{\frac{1}{n_{ij}-1}}}\\
    =\frac{10^4  \times C_{1,1} \times m_{pl}^{\frac{3n_{ij}-5}{n_{ij}-1}} \times \mu^{\frac{2}{n_{ij}-1}}}{ C_{h,k,l}^{\frac{2}{n_{ij}-1}}}
\end{align}
Instead of considering separate $n_i$ and $n_j$ for $\langle \nu_i^c \rangle$ and $\langle \nu_j^c \rangle$, we again consider $n_i = n_j = n_{ij}$ for some fractional $n_{ij}$. Following the analysis of the previous section, we find
\begin{align}
	(D^\intercal V \Lambda^{-1} V^{-1} D)_{ij}  = RM_{ij}= \frac{C_{2,1}}{m_{pl}} \langle \bar{\nu}^c_i \rangle \langle \bar{\nu}^c_j \rangle.\\
\end{align}
We will analysis the upper bound for $m_3$ in many scenarios and deduce those the rest of the neutrinos accordingly. For convenience, we let $m_1=\frac{1}{k} m_3$ and $m_2=\frac{1}{h} m_3$. Then we get
\begin{align}
    \frac{1}{m_3} \Big[(D^\intercal V)_{i3}  (V^{-1} D)_{3j} + h (D^\intercal V)_{i2}  (V^{-1} D)_{2j} + k (D^\intercal V)_{i1}  (V^{-1} D)_{1j} \Big] 
    =RM_{ij}
\end{align}
which implies
\begin{align}\label{m_3 estimate}
    m_3 = \frac{(D^\intercal V)_{i3}  (V^{-1} D)_{3j} + h (D^\intercal V)_{i2}  (V^{-1} D)_{2j} + k (D^\intercal V)_{i1}  (V^{-1} D)_{1j}}{RM_{ij}}.
\end{align}
Now, before we use inequality in Eq.~\ref{RM upper bound} to estimate the bound, we should consider specific limiting cases and get the best bound. 

First, we consider all masses are of the same order, i.e, $k=h=\mathcal{O}(1)$. Consider $i=j=2$ the numerator is $\mathcal{O}(10)$, and the upper bound is
\begin{align}
    m_3 \approx \frac{10}{RM_{11}} \le \frac{ 10 C_{h,k,l}^{\frac{2}{n_{2}-1}}}{10^4  \times C_{1,1} \times m_{pl}^{\frac{3n_{2}-5}{n_{2}-1}} \times \mu^{\frac{2}{n_{2}-1}}}<10^{-12}
\end{align}
for all $n_2 \ge 2$ where we use $C \in [10^{-15}, 1]$. As the largest mass splitting is $10^{-10.5}$ GeV, it rules out the possibility of equal magnitude for neutrino masses.

A second case is when $m_1$ and $m_2$ are of the same magnitude but much smaller then $m_3$. Then $m_3$ will be approximately the mass splitting which is $10^{-10.5}$ GeV and $h \approx k \gg 1$. However, due to the smaller mass splitting $10^{-11.5}$ GeV, we need  $m_1 \approx m_2 \gg 10^{-11.5}$ which implies $h \approx k \ll 10$. If we consider $(i,j)=(1,2)$, we find

\begin{align}
    m_3 \le \frac{10}{RM_{12}} \le \frac{ 10 C_{h,k,l}^{\frac{2}{n_{12}-1}}}{10^4  \times C_{1,1} \times m_{pl}^{\frac{3n_{2}-5}{n_{12}-1}} \times \mu^{\frac{2}{n_{12}-1}}}<10^{-12}.
\end{align}
 Thus, $m_3$ fails to satisfy the mass splitting constraint in this case. 

Finally, when $m_1 \ll m_2, m_3$, the magnitude of each entry in $RM_{ij}$ is determined by the magnitude of $m_1$. The estimate in Eq.~\ref{m_3 estimate} will be dominated by $k$ and provide an upper bound larger than the mass splitting. Hence this is a viable case that agrees with experimental observation. Nonetheless, as mentioned in Sec.~\ref{majorana}, $m_3$ cannot be massless in this model. Thus, in general, we predict the lightest neutrino to be massive but light comparing the other two. This implies
\begin{align}
    & m_3 \approx 0.05 \text{ eV} & m_2 \approx 0.009 \text{ eV}
\end{align}

\subsection{Inverted Hierarchy Analysis}
We can carry out a similar analysis for the inverted hierarchy of left handed neutrino masses. Notice that the oscillations for each label $i$ for $m_i$ do not change. The only thing we need to modify is the diagonal mass matrix
\begin{align}
  	\Lambda \equiv \diag( m_1, m_2, m_3)= \diag( x, \sqrt{x^2 + \Delta m_{21}^2  }, \sqrt{x^2-\Delta m_{31}^2},).
\end{align}
As $m_2$ is the largest, we will mimic the previous analysis as $m_1 = \frac{1}{h} m_2$ and $m_3 = \frac{1}{k} m_2$ and end up with
\begin{align}
     m_2 = \frac{k( D^\intercal V)_{i3}  (V^{-1} D)_{3j} +  (D^\intercal V)_{i2}  (V^{-1} D)_{2j} + h (D^\intercal V)_{i1}  (V^{-1} D)_{1j}}{RM_{ij}}.
\end{align}
First, we consider all masses are of the same order, i.e,$k=h=O(1)$.  Then, consider $i=j= 2$. We arrive at the same conclusion of $m_2 < 10^{-12}$ which fails to satisfy the mass splitting constraint. Unlike the normal hierarchy, the second case where $m_1 \approx m_3 \ll m_2$ is not possible with inverted hierarchy. As the large mass splitting $\Delta m_{32}^2$  requires $m_1 \approx m_3  > 10^{10.5}$, the small mass splitting $\Delta m_{12}^2 \ll \Delta m_{32}^2$ will imply $m_2 \approx m_1$ . Again, we arrive at the conclusion the lightest left handed neutrino, in this case $m_3$, is light compared to the other two. This implies
\begin{align}
    & m_1 \approx m_2 \approx 0.05 \text{ eV} 
\end{align}

The results from both hierarchies are consistent with the current knowledge of light neutrinos, for instance, the work of Gonzalo et al \cite{gonzalo2021ads}. 
\section{Ratios of Dirac and Majorana contributions}\label{Dirac-Majo ratio}
It is also important to study the percentage of Dirac and Majorana components in the three light neutrinos. From Eq.~\ref{See-Saw}, we have
\begin{align}
    E = D^{-1}V \Lambda
\end{align}
Following the previous discussion, for the normal hierarchy, we use 
\begin{align}
    \Lambda = \diag(x, 0.009 \times 10^{-9}, 0.05 \times 10^{-9})
\end{align}
where $x$ is nonzero and smaller then $10^{-12}$ GeV. Then, varying the value of $x$, element of $E$ is of order at most $\mathcal{O}(10^{-6})$. Recall that $V$ is chosen to be orthonormal and hence of order $\mathcal{O}(1)$. Thus, the ratio of Majorana components to Dirac components is less than $10^{-6}$.

Similarly, we consider inverted hierarchy with 
\begin{align}
    \Lambda = \diag(0.05 \times 10^{-9}, 0.005 \times 10^{-9}, x)
\end{align}
where $x$ is nonzero and smaller then $10^{-12}$ GeV. Although the Majorana components look slightly different, we arrive at the same conclusion that the ratio of Majorana components to Dirac components is less than $10^{-6}$.

This would predict light neutrinos are mostly Dirac type. Theoretically, this is consistent with the current works, such as that of Gonzalo et al \cite{gonzalo2021ads}. Experimentally, that implies they behave as four component Dirac spinors, and the double beta decay would be small \cite{Oberauer}. There are several experiments in progress of testing this \cite{Bilenky_2015}. 

\section{Heavy neutrino mass}\label{heavy mass}
We can also extract some information about heavy neutrinos by considering the eigenvector equations similar to Eq.~\ref{eigen}
\begin{align}\label{eigen2}
    \begin{pmatrix} 
	& 0 &D\\
	&D^\intercal & RM
	\end{pmatrix}
	\begin{pmatrix}
	&V'\\
	&E'
	\end{pmatrix}=\begin{pmatrix}
	&V'\\
	&E'
	\end{pmatrix} \Lambda'
\end{align}
where $\Lambda' $ is the diagonal mass matrix of the heavy neutrinos. In contrast with light neutrinos, We expect $V'$ to be small compared to $E'$. Similarly to light neutrino case, we can pick $E$ to be orthonormal.This would imply
\begin{align}
    D E' = V' \Lambda'\\
    D^\intercal V' + RM E' = E' \Lambda'
\end{align}
As both $D$ and $V'$ are small compared to $RM$ and $E'$ respectively, we have the estimation
\begin{align}
    RM E' \approx E' \Lambda'
\end{align}
or
\begin{align}
    E'^{-1} RM E' \approx  \Lambda'.
\end{align}
As $E'$ is orthonormal, we conclude that $\Lambda'$ is approximately the diagonalized matrix of $RM$. This means the lower bound for the heaviest eigenvalue is
\begin{align}
  \lambda_{max} \ge \frac{\tr(RM)}{3} \gtrsim 10^{14} \hspace{1 cm} GeV
\end{align}
Using this, we can estimate the upper bound for the lightest of the heavy neutrinos. 
\begin{align}
   \prod_{i=1,2,3} \lambda^{heavy}_i  =\det{RM}.
\end{align}
Hence,
\begin{align}
    \lambda^{heavy}_{min} \le \big( \frac{\det (RM)}{\lambda_{max} } \big)^{\frac{1}{2}}.
\end{align}
$\det (RM)$ is inversely proportional to the mass of the lightest neutrino, so in general $\det (RM)$ is not bounded above when the lightest neutrino becomes lighter and lighter. On the other hand, in the heaviest case, the lightest neutrino is about $10^{-11.5}$ GeV, $\det (RM)$  is about $\mathcal{O}(10^{39})$. Then, the upper bound for the lightest heavy neutrino is
\begin{align}
    \lambda^{heavy}_{min} \le 10^{12.5} \hspace{1 cm} GeV
\end{align}
\section{Conclusion}
In this paper, our primary goal is to analyze the mass matrix of neutrinos using the result from a localized model of M theory compactified on $G_2$ manifold with resolved $E_8$ singularity \cite{Gonzalez:2020nyy}. We learn in this work that the neutrinos originate in the need for the full content of the representations of the resolved $E_8$ singularity. Similar to the work of Acharya et al \cite{Acharya:2016kep}, there are two main contributions: pure neutrino mixing, and neutralinos and higginos mixing with neutrinos. We argue that the former is more significant and therefore the focus of the paper.

Dirac terms of the neutrino mass matrix are explicitly computed from the moduli of the localized model on $G_2$ manifold. We computed the contribution on the cubic level. The texture of the neutrino masses is highly hierarchical as a result of the correlation to hierarchy from the up-type quark. From experimental data of the mixing angles and mass splittings, assuming the normal ordering, we can use the Dirac terms to compute the Majorana mass matrix as a function of the lightest neutrino mass.

The Kolda-Martin mechanism is the main theoretical tool to generate Majorana terms in this paper. In this picture, the right handed neutrinos ( and their anti particles) get VEVs and generate Majorana masses through quadric terms. The VEVs along with the Dirac terms and experimental data oscillation angles create an upper bound for the masses of left handed neutrinos. Considering this upper bound in both scenarios of normal and inverted hierarchies, we conclude that the last neutrino should always be light comparing the the other two families regardless of the choice of hierarchy. However, the model and the computed Dirac terms generally forbid the lightest neutrino to be massless. The very light mass of the one of the neutrinos implies that the other two left handed neutrinos have masses about $0.05 \text{ eV}$  and $0.009  \text{ eV}$  ($0.05 \text{ eV} $ and $0.05 \text{ eV} $))  for normal (inverted) hierarchy. Moreover, the ratio of Majorana components to Dirac components is less than $10^{-6}$ for the three light neutrino in both hierarchy scenarios. This leads to the prediction that in both hierarchies, the light neutrinos are mostly Dirac type. Hence neutrinoless double-beta decay will be small. This is a testable result in a near future. On the other hands, we achieve some restriction on heavy neutrinos. The bounds are not stringent enough to make a testable prediction. 

For future work, we expect more predictive results when we understand better about the contributions from the global structures which determine all the coefficients, including those being tunable in our local theory. Locally, it is also intriguing to explore the uniqueness of the solution. If other solutions exist, it is interesting to see the implication on the physics, especially the neutrinos. As our work can be repeated for other solutions in a relatively straightforward way, it is inviting to examine a large class of solutions using bigger computational power.

\bibliography{citations}  
  
\end{document}